%% file: SJMD.tex
\documentclass[conference]{IEEEtran}
\IEEEoverridecommandlockouts
\usepackage{cite}
\usepackage{amsmath,amssymb,amsfonts}
\usepackage{algorithm}
\usepackage{algpseudocode}
\usepackage{graphicx}
\usepackage{textcomp}
\usepackage{xcolor}
\usepackage{multirow} 
\usepackage{hyperref}

\def\BibTeX{{\rm B\kern-.05em{\sc i\kern-.025em b}\kern-.08em
    T\kern-.1667em\lower.7ex\hbox{E}\kern-.125emX}}
\begin{document}

\title{Successive Jump and Mode Decomposition\\
% {\footnotesize \textsuperscript{*}Note: Sub-titles are not captured for https://ieeexplore.ieee.org  and
\thanks{This work was supported by IRRAS USA Inc., San Diego, CA 92130 US.}
}

\author{\IEEEauthorblockN{1\textsuperscript{st} Mojtaba Nazari}
\IEEEauthorblockA{\textit{Electrical and Computer Engineering} \\
\textit{Aarhus University}, Aarhus, Denmark\\
e-mail: m.nazari@ece.au.dk}
\and
\IEEEauthorblockN{2\textsuperscript{nd} Anders Rosendal Korshøj}
\IEEEauthorblockA{\textit{1: Neurosurgery, Aarhus University Hospital} \\
\textit{2: Clinical Medicine, Aarhus University}\\
Aarhus, Denmark \\
e-mail: anders.r.korshoj@clin.au.dk}
\and
\IEEEauthorblockN{3\textsuperscript{rd} Naveed ur Rehman}
\IEEEauthorblockA{\textit{Electrical and Computer Engineering} \\
\textit{Aarhus University}, Aarhus, Denmark\\
e-mail: naveed.rehman@ece.au.dk}
}

\maketitle

\begin{abstract}

We propose fully data-driven variational methods, termed successive jump and mode decomposition (SJMD) and its multivariate extension, successive multivariate jump and mode decomposition (SMJMD), for successively decomposing nonstationary signals into amplitude- and frequency-modulated (AM-FM) oscillations and jump components. Unlike existing methods that treat oscillatory modes and jump discontinuities separately and often require prior knowledge of the number of components ($K$)—which is difficult to obtain in practice—our approaches employ successive optimization-based schemes that jointly handle AM-FM oscillations and jump discontinuities without the need to predefine $K$. Empirical evaluations on synthetic and real-world datasets demonstrate that the proposed algorithms offer superior accuracy and computational efficiency compared to state-of-the-art methods.

\end{abstract}


\section{Introduction}

Nonstationary signal decomposition methods are crucial for extracting inherent bandlimited components from complex signals. Classically, Empirical Mode Decomposition (EMD) and Variational Mode Decomposition (VMD), focus on amplitude- and frequency-modulated (AM-FM) components \cite{Huang98,Dragomiretskiy14} and are widely used in various applications \cite{akbari2022identification, gupta2015baseline}. However, these methods primarily target oscillatory components and struggle in scenarios with trends, oscillations, and jumps, often mistakenly modeling jumps as oscillations \cite{stallone2020new}. Filtering techniques also face challenges in separating jumps due to their wideband spectral impact, complicating the isolation of meaningful signal components.

To address these challenges, a class of algorithms has been developed to handle jumps and noise more effectively \cite{kim2009ell_1, storath2014jump, cicone2022jot}. For instance, \cite{kim2009ell_1} introduces the $\ell_1$ trend filtering method, a variant of Hodrick–Prescott filtering, for estimating piecewise linear trends. Inverse Potts energy functionals \cite{storath2014jump} include further advancements that recover jump-sparse and sparse signals using non-convex sparse regularizers (i.e., $\ell_0$ pseudonorm-based penalty). Researchers proposed the Jumps, Oscillations, and Trends (JOT) model to decompose signals into three components \cite{cicone2022jot,Huska2023}. The JOT method effectively separates signal components without artifacts, making it suitable for preprocessing. Despite this, JOT treats all oscillations as a single component, struggling to obtain AM-FM oscillatory components. Additionally, JOT's two-stage process has many input parameters, making tuning complex, especially when combined with other AM-FM mode decomposition methods such as SVMD.

Jump Plus AM-FM Mode Decomposition (JMD) was recently proposed for the joint extraction of jump discontinuities and amplitude- and frequency-modulated (AM-FM) modes \cite{nazari2025jump}. JMD formulates an optimization problem that is simultaneously solved for $K$ AM-FM modes and a jump component using two distinct priors: one for narrowband constraints on AM-FM modes and another for minimizing jump discontinuities. While JMD provides a robust alternating direction method of multipliers (ADMM)-based solution with fewer input parameters and a faster convergence rate than existing methods, it is computationally intensive and requires prior knowledge of $K$.

To overcome these limitations, we introduce Successive Jump and Mode Decomposition (SJMD) and its multivariate extension Successive Multivariate Jump and Mode Decomposition (SMJMD), which decompose a nonstationary signal into a jump component $v(t)$ and $K$ oscillatory (AM-FM) modes $u_k(t)$ using a successive approach. Their optimization framework incorporate three priors: enforcing band-limited AM-FM modes, maximizing separation between residuals and extracted modes, and minimizing a function of the jump component’s derivative. Unlike JMD, our methods do not require prior knowledge of $K$, offering a more flexible and adaptive decomposition with a higher likelihood of convergence, reduced computational complexity, and enhanced robustness against noise.

The remainder of this paper is organized as follows: Section II presents the proposed formulation and methodology, Section III details experimental evaluations on synthetic and real-world signals, and Section IV provides concluding remarks.

\section{Successive Jump and Mode Decomposition}

In this section, we propose a variational SJMD model to extract AM-FM modes $u_k(t)$ and the jump component $v(t)$ from the input signal $s(t)$, based on the model: $s(t)=v(t) + u_{k}(t) + r(t)$, where $r=(\sum u_{i\neq k}(t) + n(t))$ refers to the residual signal (including all yet-to-be-extracted modes). The aim is to extract a single mode $u_{k}$ at each iteration rather than concurrently decomposing all $K$ modes, then removing the effect of previously obtained modes by subtracting them from the signal one after the other \cite{nazari2024multiscale}. The successive procedure helps alleviate the need to select a precise value for the parameter $K$ \cite{nazari2020successive}. Therefore, our method decomposes the signal by: i) extracting jumps using a sparsity-inducing regularizer, and ii) extracting the oscillatory components via successive minimization of a bandwidth-related term, effectively separating discontinuities from oscillations.

\subsection{Optimization Formulation}
In this framework, the assumption is that each oscillatory mode is predominantly localized around a central frequency \cite{Dragomiretskiy14} while the discontinuities are wide-band components on the spectrum. Considering these two opposite features, incorporating two different terms in our formulation is required. To obtain the AM-FM oscillatory components from the data, we use the optimization formulation of SVMD that aims to minimize both the bandwidth of the AM-FM component $u_{k}$ and the overlap between $u_{k}$ and $ r$ successively until $k=K$:
\begin{equation}
\small
\begin{aligned}
% \tilde{r}_{n}^{(m),i+1}=
J_1= 
\footnotesize \Big\Vert\partial_t\Big[{{u_k}_+}(t) e^{-j\omega_{k}t}\Big]\Big\Vert^2_2
+\Big\Vert h_k(t)*{r}(t)\Big\Vert^2_2,
\label{vmdcost} 
\end{aligned}
\end{equation}

\noindent where $\partial_t$ denotes the partial derivative, $u_{k+}(t) = u_k(t) + j \mathcal{H} u_k(t) = a_k(t) e^{j \phi_k(t)}$ (and $\hat{u}_{k+}(\omega) = (1 + \text{sgn}(\omega)) \hat{u}_k(\omega)$) is the analytic signal with $\mathcal{H}$ as the Hilbert transform, and $\omega_k$ represents the center frequency of each mode. The squared $l^2$-norm term estimates the bandwidth of the $k$-th mode, where minimizing the bandwidth approximates it as a nearly pure sinusoid. The second term, which is the key distinction of our method compared to JMD, ensures that the successive procedure is performed through enforcing the energy of the residual signal $r(t)$ to be minimized around the center frequency of the desired mode $\omega_k$. This is done using a filter $h_k(t)$ with a frequency response of $\hat{h}_k(\omega)=1/\alpha(\omega-\omega_{k})^2$, where $h_k(\omega) \rightarrow \infty$ if $\omega\rightarrow\omega_k$. 

To extract the jump component, we use a penalty term that penalizes the derivative of the jump component to promote a piecewise constant component \cite{cicone2022jot,nazari2025jump,Huska2023}. It aims to accurately preserve the amplitude of the piecewise-constant signal component while improving the representation of sharp discontinuities in the jump component:
\begin{equation}
\small
\begin{aligned}
% \tilde{r}_{n}^{(m),i+1}=
J_2= \int_{0}^\infty \phi (| \partial_t v(t) |; b) \quad dt,
\label{jumpcost}
\end{aligned}
\end{equation}
\noindent where $\partial_t v(t) := v'(t)$ represents the first derivative of the jump component. The non-convex sparsity-promoting penalty function $\phi(\cdot; b) : [0, +\infty) \to [0, 1]$ is defined as a piecewise-quadratic function \cite{cicone2022jot}, as follows:
\begin{align}
\centering
\phi (x; b) &= \begin{cases}
-\frac{b}{2}x^2 + \sqrt{2b} x & x \in [0,\sqrt{\frac{2}{b}}), \\
1 & x \in [\sqrt{\frac{2}{b}}, +\infty),
\end{cases} &\quad
\label{phifunc}
\end{align}

\noindent where the parameter $b$ changes the degree of non-convexity, such that for $b \xrightarrow{} 0$, the MC penalty is defined as $\phi (x; b) = |x|$ and $\phi (\cdot; b)$ tends to $l_0$-pseudonorm for $b \xrightarrow{} \infty$. This formulation has also been used successfully to extract jump components from input data \cite{cicone2022jot,nazari2025jump,Huska2023}.

In our proposed approach to obtain $v(t)$ and $u_k(t)$ using input data $f(t)$, we combine \eqref{vmdcost} and \eqref{jumpcost} within a single constrained optimization formulation as follows:
\vspace{-0mm}
\begin{equation}
\begin{split}
\small
\begin{aligned}
(&\overset{*}{u}_k,\overset{*}{\omega_{k}},\overset{*}{r},\overset{*}{v},x)\xleftarrow{}\underset{{{u}}_k,\omega_{k},r,v,x}{\text{arg min}} \mathcal{J}({{u}}_k,\omega_{k},r,v, x)  \\
&\mathcal{J}({{u}}_k,\omega_{k},r,v,x)=\Big\{\alpha\Big\Vert\partial_t\Big[{u}_{k+} (t) e^{-j\omega_{k} t}\Big]\Big\Vert^2_2 \\
& +\Big\Vert h_k(t)*{r}(t)\Big\Vert^2_2 + \beta\int_{0}^\infty \phi (|x(t)|; b) \hspace{1mm} dt \\
&+ \Big\Vert s(t) - \Big(v(t) + r(t) +  u_k(t)\Big)\Big\Vert^2_2\Big\},  \quad\quad\quad s.t. \quad x= \partial_t v.
\label{opt1} 
\end{aligned}
\end{split}
\end{equation}\vspace{-3.8mm}

\noindent where $\alpha$ and $\beta$ are balancing parameters for the first and third terms the auxiliary variable $x:= \partial_t v$ is performed as a splitting technique to solve the problem of non-differentiability of the term $\phi ( \cdot ; b)$. 
Next, we convert \eqref{opt1} into an unconstrained formulation using Lagrangian multipliers, the following augmented Lagrangian function is obtained:\vspace{-5mm}

\begin{equation}
\small
\begin{aligned}
&\mathcal{L}({{u}}_k,\omega_{k},r,v,x,\rho) = \Big\{ 
\alpha\Big\Vert\partial_t\Big[{u}_{k+} (t) e^{-j\omega_{k} t}\Big]\Big\Vert^2_2 + \Big\Vert h_k(t)* {r}(t)\Big\Vert^2_2\\
&+ \Big\Vert s(t) - \Big(v(t) + r(t) + u_k(t)\Big)\Big\Vert^2_2 + \beta \int_{0}^\infty \phi (| x(t) |; b) dt \\
&- \Big\langle \rho(t), x(t) -\partial_t v(t)\Big\rangle + \frac{\gamma}{2}\Big\Vert x(t) - \partial_t v(t)\Big\Vert^2_2\Big\},
\label{lagran}
\end{aligned}
\end{equation}\vspace{-5mm}

\noindent where $\gamma$ is the penalty scalar parameter and $\rho(t)$ is the vector of Lagrangian multipliers associated with constraint $x= \partial_t v$. The above formulation is divided into several sub-optimization problems and solved using ADMM. The sub-optimization problem to solve for $u_k$, $ r$, and $\omega_k$ can be written as:
\begin{equation}
\small
\begin{aligned}
&\mathcal{L}({{u}}_k,\omega_{k},r) = \Big\{\alpha\Big\Vert\partial_t\Big[{u}_{k+} (t) e^{-j\omega_{k} t}\Big]\Big\Vert^2_2  
 + \Big\Vert h_k(t)* {r}(t)\Big\Vert^2_2 \\
&+\Big\Vert s(t) -\Big(v(t) + r(t) + u_k(t)\Big) \Big\Vert^2_2\Big\}
\label{lagran2}.
\end{aligned}
\end{equation}
\vspace{-2mm}

As suggested in \cite{nazari2020successive}, by assuming $\hat{h}_k(\omega)=1/\alpha(\omega-\omega_{k})^2$, solving \eqref{lagran2} in the Fourier domain, taking the partial derivative w.r.t. each of $u_k$, $ r$, and $\omega_k$, and some algebraic manipulations, the associated update equations can be obtained as follows:
\begin{equation}
\small
\begin{aligned}
\hat{u}_{k}^{(i+1)} = \frac{\hat{s}(\omega) - \hat{r}^{(i)}(\omega) - \hat{v}^{(i)}(\omega)}{1+2\alpha(\omega-\omega_{k}^{(i)})^2}, 
\end{aligned}
\label{uk}
\end{equation}
\begin{equation}
\small
\omega_{k}^{(i+1)}=\frac{\int^{\infty}_0 \omega \Big|\hat{u}_{k}^{(i+1)}(\omega)\Big|^2 d\omega}{\int^{\infty}_0 \Big|\hat{u}_{k}^{(i+1)}(\omega)\Big|^2 d\omega},
\label{omega}
\end{equation}
\begin{equation}
\small
\begin{aligned}
\hat{r}^{(i+1)} =\frac{\alpha^2 (\omega - \omega_k^{(i+1)})^4 \left( \hat{s}(\omega) - \hat{u}_k^{(i+1)}(\omega) - \hat{v}^{(i)}(\omega) \right)}{1 + \alpha^2 (\omega - \omega_k^{n+1})^4}.
% \frac{\hat{s}(\omega) - \hat{v}(\omega) - \hat{u}_k(\omega) }{1+|\frac{1}{\alpha(\omega-\omega_{k})^2}|^2}.
\end{aligned}
\label{r}
\end{equation}

The sub-optimization problem to solve for $v$, $x$, and $\rho$ can be achieved from \eqref{lagran} as follows:
\vspace{-2mm}
\begin{equation}
% \small
\begin{aligned}
&\resizebox{1\hsize}{!}{$\mathcal{L}(v,x,\rho) = \beta \int_{0}^\infty \phi (| x(t) |; b) dt + \frac{\gamma}{2}\Big\Vert x(t) - \Big(\partial_t v(t) + \frac{\rho(t)}{\gamma}\Big)\Big\Vert^2_2$}\\ 
&+ \Big\Vert s(t) - 
\Big(v(t) + r(t) + u_k(t)\Big) \Big\Vert^2_2 .
\label{lagran3_0}
\end{aligned}
\end{equation}

The presence of an infinite integral in the term involving $\phi$ poses a significant challenge when attempting to solve this sub-problem in the continuous domain. However, it can be discretized and is strongly convex under conditions from \cite{huska2019convex}. Thus, by assuming that we deal with a finite signal (that is also numerically established for \eqref{lagran2} as explained in \cite{Dragomiretskiy14}), we can rewrite \eqref{lagran3_0} in the discrete form as follows: 
\vspace{-10mm}
\begin{equation}
\small
\begin{aligned}
&\mathcal{L}(v,x,\rho)= \beta \sum_{j=1}^{N} \phi (| x_j |; b) + \frac{\gamma}{2}\Big\Vert x - \Big(Dv + \frac{\rho}{\gamma}\Big)\Big\Vert^2_2 \\
&+ \Big\Vert s - \Big(v + r + u_k\Big) \Big\Vert^2_2 ,
\label{lagran3}
\end{aligned}
\end{equation}
\vspace{-3mm}

\noindent where $D$ is defined as the first-order derivative matrix as defined in \cite{nazari2025jump}. Next, as suggested in \cite{nazari2025jump,Huska2023}, by solving \eqref{lagran3}, taking the partial derivative w.r.t. $v$, and zero-initializing the vectors $x$ and $\rho$, we will have:
\vspace{-1mm}
\begin{equation}
\small
\begin{aligned}
v^{(i+1)} = (\gamma D^TD + 2I)^{-1} \Big(D^T\rho^{(i)} + \gamma D^T x^{(i)} + 2(s - r - u_k)\Big).
\label{vjump}
\end{aligned}
\end{equation}
By rewriting the sub-problem and omitting the constant terms, the $i$-th iteration of the associated update equation for $x$
can be expressed in terms of $N$ independent one-dimensional optimization problems as follows:
\vspace{-3mm}
\begin{equation}
\small
\begin{aligned}
x_j^{(i+1)} = \arg\min_{ x_j \in \mathbb{R}} \Big\{  \mu \phi(| x_j|; b) + \frac{1}{2} \lVert  x_j - \big[(Dv^{(i)})_j + \frac{\rho^{(i)}_j}{\gamma}\big] \rVert_2^2 \Big\},\\ j = 1, \ldots, N,
\label{xder}
\end{aligned}
\end{equation}

\noindent where $\mu = \frac{\beta}{\gamma}$. Readers are encouraged to read \cite{nazari2025jump,Huska2023} for more details. As demonstrated in \cite{huska2019convex}, the problems in \eqref{xder} are strongly convex if and only if $b < \frac{1}{\mu} \Rightarrow \gamma > b\beta \Rightarrow \gamma = \tau b\beta, \text{ for } \tau \in \mathbb{R}, \tau > 1$. Given this assumption, closed-form solutions for the problems stated in \eqref{xder} can be derived as:
\vspace{-2mm}
\begin{equation}
\small
\begin{aligned}
x_j^{(i+1)} =\min \big( \max \big( \frac{1}{1 - \mu b} - \frac{\frac{\mu\sqrt{2b}}{1-b\mu}}{\lvert h_j \rvert}, 0 \big), 1 \big)  h_j,
\label{xupdate}
\end{aligned}
\end{equation}
 \vspace{-3mm}

\noindent where $h_j= (Dv^{(i)})_j + \frac{\rho^{(i)}_j}{\gamma}$. The update for $\rho$ will be governed by the following equation: 
\vspace{-1mm}
\begin{equation}
\small
\begin{aligned}
\rho^{(i+1)} = \rho^{(i)} - \gamma \left( x^{(i+1)} - Dv^{(i+1)} \right).
\label{rho}
\end{aligned}
\end{equation}
\vspace{-6mm}

With all update equations at our disposal, the steps for obtaining the AM-FM modes $u_k(t)$ and jump component $v(t)$ are listed in Algorithm \ref{SJMD}. In the SJMD algorithm, we employ an \textit{increasing $\alpha$} strategy, starting with a low $\alpha$ and progressively raising it to the user-defined maximum, which accelerates convergence \cite{nazari2020successive,nazari2024multiscale}. 

\begin{algorithm}
    \caption{\bf SJMD}
    \label{SJMD}
    \small
    \textbf{Input and user-defined parameters:} $s, \alpha_{\max}, \beta, \overline{b},\tau$
    \begin{algorithmic}[1]
    \vspace{-1.5mm}
    \State  $(p, k) \gets 0, T= samples$\vspace{-1mm}
        \Repeat
        \vspace{-1.1mm}
        \State  $\epsilon \gets 10^{-7}, (\hat{u}_{k}^{(1)}, \omega_k^{(1)}, x^{(1)},v^{(1)},r^{(1)}, \rho^{(1)}, i)\gets 0,$ \\$ b = 2/{\overline{b}^2}, \quad \gamma = \tau b \beta,$ \text{generate discrete operator} $D$
        \State $k \gets k+1$\vspace{-1.1mm}
            \Repeat\vspace{-1.2mm}
                \State $p \gets p+1$\vspace{-1.2mm}
                \Repeat\vspace{-1mm}
                    \State $i \gets i+1$
                    \State $\hat{u}_{k}^{(i+1)}(\omega) \gets \textit{Using \eqref{uk}}$
                    \State $\omega_{k}^{(i+1)} \gets \textit{Using \eqref{omega}}$
                    \State $\hat{r}^{(i+1)}(\omega) \gets \textit{Using \eqref{r}}$   
                    \State $u_k \xleftarrow{IFT} \hat{u}_k(\omega)$ and $r \xleftarrow{IFT} \hat{r}(\omega)$ \textit{(to compute $v$)}
                    \State $v^{(i+1)} \gets \textit{Using \eqref{vjump}}$
                    \State $x^{(i+1)} \gets \textit{Using \eqref{xupdate}}$
                    \State $\rho^{(i+1)} \gets \rho^{(i)} - \gamma ( x^{(i+1)} - Dv^{(i+1)})$
                    \State $g^{(i+1)} = {u}_{k}^{(i+1)} + {v}^{(i+1)}$
                    \State $\hat{v}^{(i+1)}(\omega) \xleftarrow{FT} v^{(i+1)}$ \textit{(to update $\hat{u}_k(\omega)$)}
                \Until{Convergence: $ \Vert g^{(i+1)}-g^{(i)}\Vert_2^2/\Vert g^{(i)}\Vert_2^2 \leq \epsilon$}
                \State $(\hat{u}_{k}^{(1)}, \omega_k^{(1)}, x^{(1)},v^{(1)},r^{(1)}, \rho^{(1)}, i)\gets 0$
                \State $\alpha_{p+1} \gets 2\alpha_{p}$
            \Until{$\alpha_{p+1} \leq \alpha_{\max}$}
            \State $\textbf{u} \gets \textit{Store all current extracted modes as vectors in a matrix}$
        \Until$\frac{1}{T} \left( \left\|  {u}_k\right\|_2^2 - \left\|{u}_{k-1} \right\|_2^2 \right) \leq \epsilon$
    \end{algorithmic}
\end{algorithm}

\vspace{-2mm}
\subsection{Successive Multivariate Jump and Mode Decomposition }

Here we extend the multivariate SJMD method (i.e., SMJMD) for the multivariate signal model 
$\bold{s}(t) = \bold{v}(t) + \mathbf{u}_{k}(t) + \bold{r}(t)$, where $\bold{s}(t) = [s_1(t), ... , s_C(t)]$ represents a multivariate signal with $C$ data channels, and $\bold{u}_k(t)$ denotes the multivariate AM-FM decomposed modes with $C$ channels. The components $\bold{v}(t)$ and $\bold{r}(t)$ are the multivariate jump and residual, respectively, with $\bold{r}(t) = \sum \bold{u}_{i \neq k}(t) + \bold{n}(t)$, where $\sum \bold{u}_{i \neq k}$ encompasses all remaining multivariate modes.

The goal is to extract the jump components $\bold v(t)$ in all channels and the frequency aligned AM-FM modes $\bold u_k(t)$ from input signal $\bold s(t)$. To achieve this, we modify the cost function of the MJMD method proposed in \cite{nazari2025jump} as follows:

\vspace{-1mm}
\begin{equation}
\begin{split}
\small
\begin{aligned}
& (\{\overset{*}{u}_{k,c}\},\{\overset{*}{\omega}_{k}\},\{\overset{*}r_c\},\{\overset{*}v_{c}\}, \{x_c\}) \\
& \hspace{5mm}\xleftarrow{}\underset{\{{{u}}_{k,c}\},\{\omega_{k}\},\{r_{c}\},\{v_{c}\}, \{x_c\}}{\text{arg min}} \mathcal{J}(\{{{u}}_{k,c}\},\{\omega_{k}\},\{r_c\},\{v_{c}\}, \{x_c\})\notag \\
\end{aligned}
\end{split}
\end{equation}
\begin{equation}
\begin{split}
\small
\begin{aligned}
&\resizebox{0.45\hsize}{!}{$\mathcal{J}(\{{{u}}_{k,c}\},\{\omega_{k}\},\{r_{c}\},\{v_{c}\}, \{x_c\})$} =
\footnotesize \Big\{ \alpha \sum_c\Big\Vert\partial_t\Big[{u}_{k,c}^+ (t) e^{-j\omega_{k} t}\Big]\Big\Vert^2_2 \hspace{16mm}\\ 
&+\Big\Vert h_k(t)*{r_c}(t)\Big\Vert^2_2 +\beta \int_{0}^\infty \phi (| x_c(t) |; b) dt \\
&+ \sum_c \Big\Vert s_c(t) -
\Big(v_c(t) + r_c(t) + u_{k,c}(t)\Big)\Big\Vert^2_2 \Big\},   \hspace{4mm} s.t. \hspace{3mm} x_c= \partial_t v_c .
\label{mJMD} 
\end{aligned}
\end{split}
\end{equation}
\vspace{-4mm}

Similar to the procedure described for the proposed SJMD in previous subsections, the above optimization formulation for SMJMD can be converted into an unconstrained formulation using Lagrangian multipliers. The Lagrangian function is then divided into several sub-optimization problems and is similarly solved using ADMM through iteratively updating ${{u}}_{k,c},\omega_{k},r_{c},v_{c}$, $x_c$, and $\rho_c$ as our presented procedure for the SJMD method in previous subsections (i.e., in equations \eqref{lagran2} to \eqref{rho} and Algorithm \ref{SJMD}). 

The MATLAB implementation codes for SJMD and SMJMD methods are available at the following link:  \href{https://se.mathworks.com/matlabcentral/profile/authors/16833607}{https://se.mathworks.com/matlabcentral/profile/authors/16833
607}.

\subsection{Input Parameters}
Accurate selection of input parameters is crucial for the convergence of the proposed method similar to \cite{nazari2025jump}. A high value of \(\alpha\) may lead to noisy modes or convergence issues, while a low \(\alpha\) causes mode mixing. Typically, setting \(\alpha\) in the range \(10^3 - 10^5\) yields satisfactory results. The regularization parameter \(\beta\) controls the jump component and should be set to \(1/\text{(number of expected jumps)}\). Fine-tuning \(\beta\) ensures optimal performance. \(\tau > 1\) is used for initializing \(\gamma\), typically between 1.1 and 50. \(\overline{b}\), the expected minimal jump height (default 0.3), determines the threshold above which jumps are penalized, and should be chosen based on the characteristics of the signal.

\begin{figure*}
\centering
\includegraphics[width=1\textwidth]{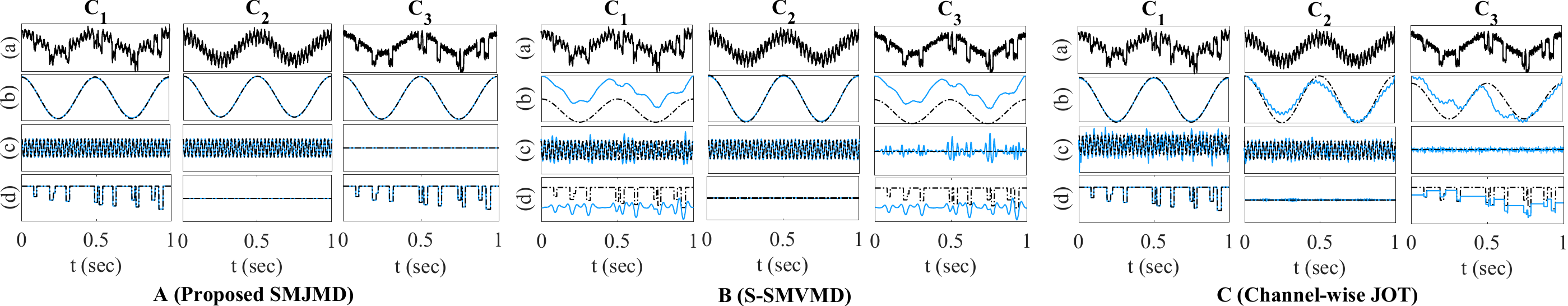}
\caption{Synthetic signal (a), and decomposed components (rows (b) - (e)) obtained from the (A) SMJMD, (B) S-SMVMD, and (C) channel-wise JOT methods. In plots (b) - (e), the solid blue line represents the extracted components, while the black dashed line indicates the original components of the signal}
\label{Synth}
\end{figure*}

\begin{figure}
\centering
\includegraphics[width= .5\textwidth]{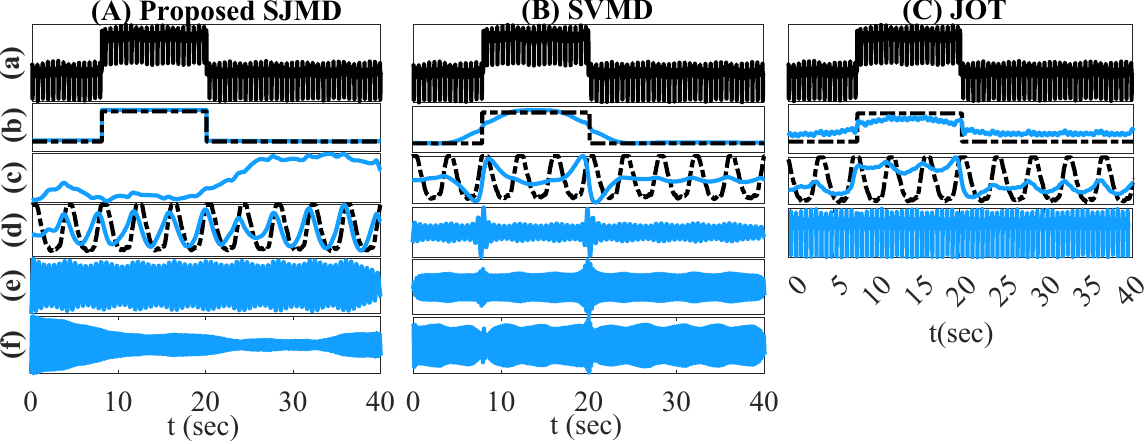}
\caption{Decomposition of the input ECG signal (record number “055m” from MIMIC database) with additive simulated jump component (a) into a jump (in row
(b) of the results for all methods) and EDR (in row (d) for our method and in row (c) for the results of
SVMD and JOT). The solid blue line refers to the extracted components and the black dashed line indicates the reference components.} 
\label{ECG}
\end{figure}

\begin{figure}
\centering
\includegraphics[width=.49\textwidth]{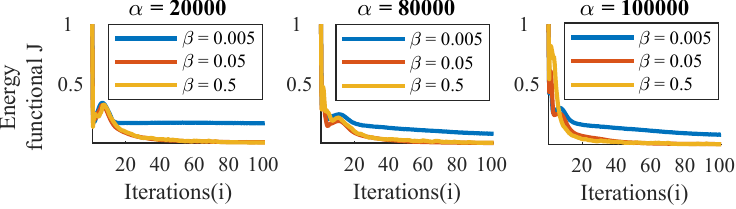}
\caption{Empirical evidence of numerical convergence of S(M)JMD for the synthetic signal in \eqref{multiSynth}.}
\label{Conv}
\end{figure}

\section{Experiments and Results}

We evaluated the performance of the SMJMD method on multivariate synthetic data, comparing it with simplified successive multivariate VMD (S-SMVMD) \cite{nazari2024multiscale} and JOT \cite{Huska2023} applied channel-wise, as JOT does not support multichannel data. Similarly, we applied SJMD to real-world signals and compared its performance with SVMD \cite{nazari2020successive} and JOT. Additionally, as S(M)JMD and (M)JMD \cite{nazari2025jump} yielded similar decomposition results, we only demonstrated the decomposition results of S(M)JMD, and instead, we assessed the computational efficiency and noise robustness of S(M)JMD against (M)JMD, where SJMD demonstrates a clear advantage. 

Since \cite{nazari2020successive} demonstrated that successive strategies require a higher $\alpha$ than concurrent approaches, we set the same $\alpha$ values for S(M)JMD, SVMD, and S-SMVMD to ensure a fair comparison. For other techniques, we used parameter values that yielded the best results. All synthetic simulations considered time discretized as $t\in [0,1]$.

\subsection{SMJMD vs. Channel-Wise JOT \cite{Huska2023} and S-SMVMD \cite{nazari2024multiscale}}
We tested the SMJMD, focusing on the method's ability to align common frequency scales across multiple data channels, which is crucial for various scientific and engineering applications involving multivariate data \cite{gupta2015baseline}. To achieve this, we generated a simple synthetic signal $\bold{s}_1(t)$ composed of three channels (i.e., $c_1, c_2$, and $c_3$), incorporating oscillatory components, discontinuities $v(t)$, and additive white Gaussian noise (AWGN) $\eta$ with zero mean and a variance of $\sigma=0.1$, as follows:
\begin{equation}
\small
\boldsymbol{s}_1(t) = 
\left\{
\begin{aligned}
c_1(t) & = \cos(2\pi 2 t) + \frac{1}{2}\cos(2\pi 40 t) + v(t) + \eta, \\
c_2(t) & = \cos(2\pi 2 t) + \frac{1}{2}\cos(2\pi 40 t) + \eta, \\
c_3(t) & = \cos(2\pi 2 t) + v(t) + \eta.
\end{aligned}
\right.
\label{multiSynth}
\end{equation}

Figure \ref{Synth} visualizes the extracted components from the proposed SMJMD method, including two oscillatory modes and one jump. Our method accurately extracted all components compared to the other methods. For instance, S-SMVMD failed to extract jumps, as its formulation was not theoretically designed to capture them. Since the jump component has a wide spectrum, significant spectral overlap with the oscillatory components was expected. This overlap caused all VMD-based methods to struggle in isolating pure modes, as they function similarly to Wiener filtering applied to the residual spectrum \cite{Dragomiretskiy14}. On the other hand, the JOT method accurately captured jumps only in $c_1$. Moreover, it failed to precisely extract the 2 Hz modes in $c_2$ and $c_3$ and the 40 Hz modes in $c_1$ and $c_2$, as illustrated in Fig. \ref{Synth}-(C). 
The input parameters of SMJMD for this experiment were set to $\alpha=80000$, $\beta=0.05$, $\overline{b}=0.9$, and $\tau=50$.

\subsection{SJMD vs. JOT \cite{Huska2023} and SVMD \cite{nazari2020successive}} 

Electrocardiography (ECG)-derived respiration (EDR) has been widely studied for sleep disorders, as respiration introduces a low-frequency component to the ECG signal. However, ECG jumps generate strong harmonics in this range, complicating EDR extraction with traditional decomposition methods. To evaluate our method's performance, we used the MIMIC Database ECG dataset with a reference respiratory signal \cite{goldberger2000physiobank}. The ECG signal was artificially contaminated with electrode motion artifacts, resulting in a -9 dB signal-to-noise ratio. Using SVMD and our proposed SJMD, we extracted 15 decomposed modes from the ECG signal, with selected modes and jumps shown in Fig. \ref{ECG}. The jump components are displayed in row (b) for all methods, while EDR is visualized in row (d) for SJMD and row (c) for SVMD and JOT. The quantitative comparison in Table \ref{comparison} (correlation coefficient (CC) and mean squared error (MSE)) demonstrates SJMD's superior performance over SVMD and JOT, effectively extracting EDR despite jumps. In contrast, JOT struggled with AM-FM EDR extraction, and SVMD failed to isolate jumps from AM-FM modes. For optimal EDR extraction, we set $\alpha = 10^5$, with $\beta = 0.9$, $\overline{b} = 0.3$, and $\tau = 50$.

\begin{table}[t]
    \setlength{\tabcolsep}{4pt}
    \renewcommand{\arraystretch}{1} % 
    \caption{Quantitative Comparison of Methods for EDR}
    \label{tab:par}
    \centering
    \begin{tabular}{ccccc}
        \hline
        \textbf{Method} & \textbf{CC of EDR} & \textbf{CC of Jump} & \textbf{MSE of EDR} & \textbf{MSE of Jump} \\ \hline
        \textbf{SJMD} & \textbf{0.9010} & \textbf{0.9991} & \textbf{0.1110} & \textbf{0.0067} \\ \hline
        SVMD & 0.1282 & 0.9057 & 0.5123 & 0.1349 \\ \hline
        JOT & 0.2861 & 0.8050 & 0.4717 & 2.1308 \\ \hline 
    \end{tabular}
    \label{comparison}
    \vspace{-.16cm}
\end{table}

\subsection{Noise Robustness and Computational Costs}

Table~\ref{CCofmodes} demonstrates the superior noise robustness of SMJMD over MJMD across different noise levels ($\sigma$) applied to the signal in \eqref{multiSynth}. Unlike MJMD, which requires fine-tuning as noise increases, SMJMD remains robust without parameter adjustments. Our observations confirm that this resilience is attributed to SMJMD's ability to store noise and distortion in the residual signal $r_c(t)$, thereby preserving the jump component with minimal distortion. This also holds for SJMD.

In addition, the elapsed times to decompose the same ECG data (mentioned in the previous subsection) were measured on a system with a 2.4 GHz Intel 11th Gen i5-1135G7 processor and 16 GB RAM. The results, presented in Table~\ref{CCofmodes}, confirm that SJMD has lower computational complexity than JMD. Moreover, the parameter $K$ was automatically determined by SJMD, whereas in JMD, it was optimally fine-tuned through a time-consuming and computationally demanding procedure.

\subsection{Convergence}

Like (M)JMD, the optimization formulations of the proposed S(M)JMD are inherently non-convex and lack formal convergence guarantees. However, extensive experiments confirm that S(M)JMD consistently converge across all tested cases. Fig.~\ref{Conv} presents the normalized convergence plots for the signal in \eqref{multiSynth}, illustrating the decay and stabilization of the objective function energy of SMJMD over $i$ iterations for varying $\alpha$ and $\beta$. This also holds for SJMD.

\begin{table}
\centering
\caption{Performance Comparison: Correlation Coefficient of Decomposed Components and Computational Burden}
\label{tab:merged_table}
\renewcommand{\arraystretch}{1.1} 
\setlength{\tabcolsep}{5pt} 
\begin{tabular}{c c c c c c c c}
\hline
\multirow{2}{*}{\textbf{Method}} & \multicolumn{4}{c}{\textbf{Correlation Coefficient (Mean ± Std. Dev.)}} \\ 
\cline{2-6}
 & & $\sigma = 0.1$ & $\sigma = 0.3$ & $\sigma = 0.6$  & \\ \hline
SMJMD & & 0.9888 ± 0.0117 & 0.9751 ± 0.0121 & 0.9546 ± 0.0165&   \\ 
MJMD  & & 0.9749 ± 0.0168 & 0.9739 ± 0.0206 & 0.9268 ± 0.0367&   \\ \hline \hline
\multirow{2}{*}{\textbf{Method}} & \multicolumn{4}{c}{\textbf{Computational Burden}} \\ 
  \cline{2-6}
  & & &\hspace{-20mm} \textbf{Number of Modes (K)}  & \textbf{Elapsed Time} \\ \hline
SJMD & & &\hspace{-20mm} (Automatically set) 15  & 61 sec. \\ 
JMD  & & &\hspace{-20mm} (Manually set) 13  & 83 sec. \\ \hline
\label{CCofmodes}
\end{tabular}
\end{table}

\section{Conclusion}

In this paper, we present novel data-driven methods, S(M)JMD, to decompose non-stationary signals into oscillatory modes and jump components. The methods combine AM-FM signal decomposition with jump extraction, capturing both sharp discontinuities and oscillatory modes, unlike SVMD, S-SMVMD and JOT. On top of that, S(M)JMD address the limitations of (M)JMD by extracting components one after the other, reducing the complexity and number of input parameters (particularly $K$), improving noise robustness. Our methods utilizes a variational optimization framework solved via ADMM that ensures convergence across various applications. Experimental results show that SJMD outperforms existing methods in both synthetic data and real-world ECG signals.

\bibliographystyle{IEEEtran}

% \bibliography{SJMD.bib}
\input{SJMD.bbl}

\end{document}

%% file: SJMD.bbl

%% file: SJMD.bbl
\begin{thebibliography}{10}
\providecommand{\url}[1]{#1}
\csname url@samestyle\endcsname
\providecommand{\newblock}{\relax}
\providecommand{\bibinfo}[2]{#2}
\providecommand{\BIBentrySTDinterwordspacing}{\spaceskip=0pt\relax}
\providecommand{\BIBentryALTinterwordstretchfactor}{4}
\providecommand{\BIBentryALTinterwordspacing}{\spaceskip=\fontdimen2\font plus
\BIBentryALTinterwordstretchfactor\fontdimen3\font minus
  \fontdimen4\font\relax}
\providecommand{\BIBforeignlanguage}[2]{{%
\expandafter\ifx\csname l@#1\endcsname\relax
\typeout{** WARNING: IEEEtran.bst: No hyphenation pattern has been}%
\typeout{** loaded for the language `#1'. Using the pattern for}%
\typeout{** the default language instead.}%
\else
\language=\csname l@#1\endcsname
\fi
#2}}
\providecommand{\BIBdecl}{\relax}
\BIBdecl

\bibitem{Huang98}
N.~E. Huang, Z.~Shen, S.~R. Long, M.~C. Wu, S.~H. H., Q.~Zheng, N.-C. Yen,
  C.~C. Tung, and H.~H. Liu, ``The empirical mode decomposition and the hilbert
  spectrum for nonlinear and non-stationary time series analysis,'' \emph{Proc.
  R. Soc. A: Math. Phys. Eng. Sci.}, vol. 454, no. 1971, pp. 903--995, 1998.

\bibitem{Dragomiretskiy14}
K.~Dragomiretskiy and D.~Zosso, ``Variational mode decomposition,'' \emph{IEEE
  Trans. Signal Process.}, vol.~62, no.~3, pp. 531--544, 2014.

\bibitem{akbari2022identification}
H.~Akbari, M.~T. Sadiq, S.~Siuly, Y.~Li, and P.~Wen, ``Identification of normal
  and depression eeg signals in variational mode decomposition domain,''
  \emph{Health Inf. Sci. Syst.}, vol.~10, no.~1, p.~24, 2022.

\bibitem{gupta2015baseline}
P.~Gupta, K.~K. Sharma, and S.~D. Joshi, ``Baseline wander removal of
  electrocardiogram signals using multivariate empirical mode decomposition,''
  \emph{Healthc. Technol. Lett.}, vol.~2, no.~6, pp. 164--166, 2015.

\bibitem{stallone2020new}
A.~Stallone, A.~Cicone, and M.~Materassi, ``New insights and best practices for
  the successful use of empirical mode decomposition, iterative filtering and
  derived algorithms,'' \emph{Sci. Rep.}, vol.~10, no.~1, p. 15161, 2020.

\bibitem{kim2009ell_1}
S.-J. Kim, K.~Koh, S.~Boyd, and D.~Gorinevsky, ``$\ell_1$ trend filtering,''
  \emph{SIAM Rev.}, vol.~51, no.~2, pp. 339--360, 2009.

\bibitem{storath2014jump}
M.~Storath, A.~Weinmann, and L.~Demaret, ``Jump-sparse and sparse recovery
  using potts functionals,'' \emph{IEEE Trans. Signal Process.}, vol.~62,
  no.~14, pp. 3654--3666, 2014.

\bibitem{cicone2022jot}
A.~Cicone, M.~Huska, S.-H. Kang, and S.~Morigi, ``Jot: a variational signal
  decomposition into jump, oscillation and trend,'' \emph{IEEE Trans. Signal
  Process.}, vol.~70, pp. 772--784, 2022.

\bibitem{Huska2023}
M.~Huska, A.~Cicone, S.~H. Kang, and S.~Morigi, ``{A Two-stage Signal
  Decomposition into Jump, Oscillation and Trend using ADMM},'' \emph{{Image
  Processing On Line}}, vol.~13, pp. 153--166, 2023,
  \url{https://doi.org/10.5201/ipol.2023.417}.

\bibitem{nazari2025jump}
M.~Nazari, A.~R. Korsh{\o}j, and N.~ur~Rehman, ``Jump plus am-fm mode
  decomposition,'' \emph{IEEE Transactions on Signal Processing}, 2025.

\bibitem{nazari2024multiscale}
M.~Nazari, A.~R. Korsh{\o}j, and N.~Rehman, ``Multiscale dynamic graph signal
  analysis,'' \emph{Signal Processing}, vol. 222, p. 109519, 2024.

\bibitem{nazari2020successive}
M.~Nazari and S.~M. Sakhaei, ``Successive variational mode decomposition,''
  \emph{Signal Process.}, vol. 174, p. 107610, 2020.

\bibitem{huska2019convex}
M.~Huska, A.~Lanza, S.~Morigi, and I.~Selesnick, ``A convex-nonconvex
  variational method for the additive decomposition of functions on surfaces,''
  \emph{Inverse Probl.}, vol.~35, no.~12, p. 124008, 2019.

\bibitem{goldberger2000physiobank}
A.~L. Goldberger, L.~A. Amaral, L.~Glass, J.~M. Hausdorff, P.~C. Ivanov, R.~G.
  Mark, J.~E. Mietus, G.~B. Moody, C.-K. Peng, and H.~E. Stanley, ``Physiobank,
  physiotoolkit, and physionet: components of a new research resource for
  complex physiologic signals,'' \emph{Circulation}, vol. 101, no.~23, pp.
  e215--e220, 2000.

\end{thebibliography}
